\documentstyle[12pt,epsf]{article}
\makeatletter
\def\eqalign#1{\null\,\vcenter{\openup\jot\m@th
  \ialign{\strut\hfil$\displaystyle{##}$&$\displaystyle{{}##}$\hfil
      \crcr#1\crcr}}\,}
\def\eqalignleft#1{\null\,\vcenter{\openup\jot\m@th
  \ialign{\strut$\displaystyle{##}$\hfil&$\displaystyle{{}##}$\hfil
      \crcr#1\crcr}}\,}
\makeatother
\def\M{{\cal M}}

\def\R{{\cal R}}
\begin{document}

\begin{center}
{\large\bf Astrophysical Sources of Stochastic Gravitational Radiation\\
in the Universe}\\
\vskip\baselineskip
\vskip\baselineskip

K.A.Postnov\\
\vskip\baselineskip

{\it Sternberg Astronomical Institute, Moscow
State University,\\
 Moscow 119899, Russia}
\end{center}

\begin{abstract}
Stochastic gravitational waves (GW) associated with unresolved
astrophysical sources at frequency bands of the ongoing GW
interferometers LIGO/VIRGO and LISA are studied.  We show that GW
noise from rotating galactic neutron stars with low magnetic fields
may reach the advanced LIGO sensitivity level at frequency $f\sim 100$
Hz.  Within LISA frequency band ($10^{-4}-10^{-1}$ Hz), the GW
background from galactic binary stars is shown to mainly contribute up
to a frequency of $3\times 10^{-2}$ Hz, depending on the galactic rate
of binary white dwarf mergers. To be detectable by LISA, relic GW
backgrounds should be as high as $\Omega_{GW}h_{100}^2>10^{-8}$ at
$10^{-2}$ Hz.

\end{abstract}  

\section{Introduction}

In a few years, with the completion of construction of gravitational
wave detectors of high sensitivity, a new window into the Universe
will be open (see Thorne 1995, Schutz 1996 for a recent review of
gravitaitonal wave astronomy).  In this connection, extensive studies
of possible sources of gravitationa radiation are now being conducted.
The most promising targets for the initial laser interferometers with
the rms sensitivity level $h_{rms}\approx 10^{-21}$ at $f=100$ Hz are
coalescing binary neutron stars and/or black holes, which can be
observable from distances up to 100 Mpc (Lipunov, Postnov \& Prokhorov
1997).

Stars are most numerous baryonic objects in the Universe ($\sim
10^{11}$ within the Galaxy, $\sim 10^{21}$ within the Hubble radius
$R_H\sim 3000h_{100}^{-1}$ Mpc, where $h_{100}=H_0/100$ km/s/Mpc is
the present value of the Hubble constant).  Among them the most
significant sources of GW are rotating triaxial neutron stars (NS) and
binary stars. Only a small fraction of galactic NS (about 700) is
observed as radiopulsars, so when trying to search for GW from them we
have an advantage of knowing the precise spin period and the position
on the sky. The same argument relates to the binary stars with known
orbital periods. But the vast majority of them are unresolved sources
and will form a stochastic background.
A stochastic GW background is commonly measured in terms of the energy
density per logarithmic frequency interval related to the critical
energy density to close the Universe, $\Omega_{GW}=dE_{GW}/d\ln
f/\rho_{cr} c^2$ ($\rho_{cr}=3H_0^2/8\pi G\approx 1.9\times 10^{-29}
h_{100}^2$ g cm$^{-3}$ where $c$ is the speed of light).  
For comparison with dimensionless detector's sensitivity $h$, 
one commonly uses the equivalent characteristic strain
$h_c(f)=(1/2\pi)(H_0/f)\Omega_{GW}^{1/2}$.    

Being interesting by themselves, astrophysical backgrounds, however, are
viewed as a noise burying a possible cosmological gravitational wave
background (CGWB), which bears the unique imprint of physical
processes occurring at the very early (near-Plankian) age of the
Universe (see e.g. Grishchuk 1988 for a review).
Exact value of CGWB is still very controversial (see Grishchuk 1996
for fresh estimates). What is more reliable (however, not completely
parameter-free) are GWBs formed by known astrophysical sources (old
neutron stars, binary stars), and here we address the question how
much astrophysical GW noises contribute at LIGO/VIRGO (10-1000 Hz) and
LISA ($10^{-4}-10^{-1}$ Hz) frequencies.

\section{GW noise from sources with changing frequency}

To calculate GW noise produced by some unresolved sources we need to
know the number of sources per logarithmic frequency interval.  At the
first glance, this would require knowing precise formation and
evolution of sources. However, when only GW carries away angular
momentum from the emitting source, the problem becomes very simple and
physically clear.

GW energy loss leads to changing the frequency $\omega=2\pi f$ of
emitting objects. In the case of a rotating triaxial body, the
positive rotational energy $E_{rot}=I\omega_{rot}^2/2$ (here $I$ is
the moment of inertia) is being lost and the spin frequency (hence, GW
freqency) decreases.  In contrast, in the case of a binary star with
masses of the components $M_1$ AND $m_2$ in a circular orbit ofr adius
$a$ the negative orbital energy, $E_{orb}= -M_1M_2/2a\sim \M c^2
f_{orb}^{2/3}$ (here $\M=(M_1+M_2)^{2/5}(M_1M_2)^{3/5}$ is the
so-called ``chirp mass'' of the system) is being lost and the orbital
frequency increases.

To a very good approximation, the conditions of star formation and
evolution of astrophysical objects in our Galaxy may be viewed as
stationary. This is true at least for last 5 billion years.  Let the
formation rate of GW sources be $\R$. For example, the mean formation
rate of massive stars ($>$10~M$_\odot$ to produce NS) is about 1 per
30 years.

%Below we shall normalize all
%calcualtions to this rate, $\R_{30}\equiv 1/30\,$yr$^{-1} \approx
%10^{-9}$ s$^{-1}$.

The stationarity implies that the number of sources per unit
logarithmic frequency interval is
\begin{equation}
dN/d\ln f \equiv N(f) = \R \times (f/\dot f)\,.
\label{contin}
\end{equation}
The total energy emitted in GW per second per unit logarithmic
frequency interval at $f$ by all such sources in the galaxy is
\begin{equation}
dE_{GW}/(dt\,d\ln f) \equiv L(f)_{GW}= \widetilde L(f)_{GW} N(f) =
\widetilde L(f)_{GW} \R \times (f/\dot f)\,,
\end{equation}
where $\widetilde L(f)_{GW}$ is the GW luminosity of the typical
source at frequency $f$ ($\widetilde L(f)_{GW}\propto f^6$ for
non-axisymmetric neutron stars, $\widetilde L(f)_{GW}\propto f^{10/3}$
for binary systems).

Finally, for an isotropic background we have
\begin{equation}
\Omega_{GW}(f)\rho_{cr} c^2 = L(f)_{GW}/(4\pi c \langle r \rangle^2)
\label{omega}
\end{equation}
where $\langle r \rangle$ is the inverse-square average distance to
the typical source.  Strictly speaking, this distance (as well as the
binary chirp mass $\cal M$ and moment of inertia $I$ of NS) may be a
function of frequency since the binaries characterized by different
$\cal M$ may be differently distributed in the galaxy. We are highly
ignorant about the real distribution of old NS and binaries in the
galaxy, but taking the mean photometric distance for a spheroidal
distribution in the form
$dN\propto \exp[-r/r_0]\,\exp[-(z/z_0)^2]$ 
($r$ is the radial
distance to the galactic center and $z$ the hight above the galactic
plane) with $r_0=5$ kpc and $z_0=4.2$ kpc with $\langle r \rangle
\approx 7.89$ kpc is sufficient for our purposes.

For the cases considered the energy reservoir radiated in GW is either
rotational energy (neutron stars) or orbital energy (binary systems),
both depending as some power of the corresponding frequency: $E\propto
f^\alpha$, $\alpha_{NS}=2$, $\alpha_{bs}=2/3$. Hence the frequecy
change $(f/\dot f)$ may be found from the equation
$
dE/dt=\alpha E (\dot f/f)
$
By energy conservation law 
$
dE/dt=(dE/dt)_{GW}+(dE/dt)_{EM}+(dE/dt)_{\ldots}
$
where index EM stands for elecromagnetic losses and $\ldots$ means
other possible losses of energy. Finally, we obtain
\begin{equation}
(f/\dot f)= \alpha E / ((dE/dt)_{GW}+(dE/dt)_{EM}+(dE/dt)_{\ldots})
\end{equation}
and 
\begin{equation}
L(f)_{GW}= \R \alpha \widetilde E \frac{1}{
1+\frac{(dE/dt)_{EM}}{(dE/dt)_{GW}}+
\frac{(dE/dt)_{\ldots}}{(dE/dt)_{GW}}}
\end{equation}
The remarkable result is that if GW is the dominant source of energy
removal, the resulting GW stochastic background depends only on the
source formation rate:
\begin{equation}
\Omega_{GW}(f)\rho_{cr} c^2 = \R \alpha \widetilde E /(4\pi c \langle
r \rangle^2)
\label{omega_R}
\end{equation}

\section{GWB from old neutron stars at LIGO/VIRGO frequencies} 

Spin evolution of rotating non-axisymmetric NS with ellipticity 
$\epsilon$ and magnetic moment $\mu$ may be driven by GW or elecgtromagnetic
losses. 
The condition that a stochastic signal appears within the detector
band deoends on the rate of the frequency change. The upper frequency
of the stochastic background for pure electromagnetic energy losses is
$ f^{EM}_0\approx
10^3(\hbox{Hz})\R_{30}^{1/2}I_{45}^{1/2}\mu_{30}^{-1} $ where
$\mu_{30}=\mu/(10^{30}\hbox{G cm}^3)$ is NS magnetic moment.  For pure
GW losses this upper frequency is $ f^{GW}_0\approx 1.4\times
10^4(\hbox{Hz}) \R_{30}^{1/4}I_{45}^{-1/4}\epsilon_{-6}^{-1/2} $ where
$\epsilon_{-6}=\epsilon/10^{6}$ is the NS ellipticity.

For plausible values of the NS magnetic fields
($\mu_{30}=10^{-4}$--$10^2$) and ellipticities
($\epsilon_{-6}=10^{-3}$--$10^2$), at any frequency $<10^3$ Hz we
deal with stochastic backgrounds from galactic NS.  Physically, this
is due to the inability of old NS to leave frequency interval
$\Delta\omega \sim \omega$ during the typical time between consecutive
supernova explosions. 

For purely GR-driven NS spin-down the resulting spectrum is
independent of the unknown value of $\epsilon$ in the NS population.
Any additional braking mechanism always lowers the resulting signal.
Taking typical values $I=10^{45}$ g cm$^2$, $\R=1/30$
yr$^{-1}$ we obtain from Eq. (\ref{omega_R})
\begin{equation}
\Omega_{NS}\approx 10^{-7}\R_{30}^{1/2}
I_{45}(f/100 Hz)^2h_{100}^{-2}(r/10 kpc)^{-2}
\end{equation}
or in terms of $h_c$
\begin{equation}
h_c \approx \frac{1}{\tilde r} \sqrt{GI\R/c^3}\,
\approx 10^{-24}\left(\frac{10\hbox{kpc}}{\tilde r}\right)
\R_{30}^{1/2}I^{1/2}_{45}
\label{h_lim}
\end{equation}
(here we assumed the characteristic distance to NS population of order
10~kpc). Remarkably, 
this limit does not depend on frequency.
The GR background of such strength could be detected by the
advanced LIGO/VIRGO interferometers in one year integration (Thorne
1987; Giazotto 1997).

For realistic NS parameters the ratio of electromagnetic to GW
losses $x=\dot E_{EM}/\dot E_{GW}$ is
\begin{equation}
x\approx 4000 \mu_{30}^2\epsilon_{-6}^{-2}
\left(\frac{100 \hbox{Hz}}{f}\right)^2
\label{x}
\end{equation}
and electormagnetic losses becomes insignificant ($x\ll 1$)
only at high frequencies 
$
f >f_{cr}\approx 6.3(\hbox{kHz})\, \frac{\mu_{30}}{\epsilon_{-6}}
$
If we would take $\epsilon_{-6}=10^{-3}$
and $\mu_{30}=10^{-4}$ as in millisecond pulsars, we would
obtain $f_{cr}\approx 630$ Hz, however millisecond
pulsars are spun up by accretion in binary systems
and are not considered here. 

Therefore, for realistic NS we must consider the 
case $x\gg 1$. 
Then the stochastic background from old NS becomes
\begin{equation}
h_c(f)\approx 5\times 10^{-28}
\left(\frac{10\hbox{kpc}}{\tilde r}\right)
\R_{30}^{1/2}I^{1/2}_{45}\epsilon_{-6}\mu_{30}^{-1}f
\label{h(nu)_em}
\end{equation}
and is lis below even advanced LIGO sensitivity at $f\sim 100$ Hz. 

We have shown that if the NS form ellipticity is present, the
stochastic GR background produced by old NS population is naturally
formed due to NS rotation braking.  In the limiting case when only GR
angular momentum loss causes NS spin-down, this background is {\it
independent\/} on both exact value of the NS form ellipticity
$\epsilon$ and frequency and can be detected by advanced LIGO/VIRGO
interferometers.  In reality, the magnetic field of NS causes more
effective electromagnetic energy loss: to be insignificant, the
magnetic field of a NS should be less than (see Eq. (\ref{x}))
$
\mu < 1.5\times 10^{26} (\hbox{G cm}^3) \epsilon_{-6}\,f
$

According to Urpin \& Muslimov (1992), the magnetic field can decay
very fastly provided that the field was initially concentrated in the
outer crust layers with the density $<10^{10}- 10^{11}$ g cm$^{-3}$,
and such very low magnetic field for old NS may be possible.  In the
limiting case that the NS magnetic field does not decay at all (for
example, if only accretion-induced field decay is possible in binary
systems (Bisnovatyi--Kogan \& Komberg 1974)), old NS should lose their
energy through electromagnetic losses and be very slow rotators with
periods of about a few seconds. Then the initial magnetic field
distribution becomes crucial. If it is centered at $\sim 10^{12}$ G
(as implied by radipulsar $P$--$\dot P$ measurements), we have little
chances to detect the old NS population at 10--100~Hz frequency band
unless close mean distances ($<$10~kpc) are assumed (Giazotto et
al. 1997). However, if nature prefers a scale-free law (like
$f(\mu)\propto 1/\mu$), the fraction of low-field NS could amount to a
few $10\%$ and they can contribute to the frequency-independent GR
background. Then Eq. (\ref{h_lim}) implies that such a background can
be detected by the advanced LIGO/VIRGO interferometer in the frequency
band 10--1000~Hz in one-year integration even if the formation rate of
such NS is as small as 1 per 300 years and the characteristic distance
to them is 100~kpc.

\section{GW noise from unresolved binary stars at LISA frequencies}

Inside LISA frequency range, $10^{-4}-10^{-1}$ Hz, only coalescing
binary white dwarfs (WD) and binary neutron stars contribute. Even if
binary neutron stars coalesce at a rate of 1/10000 yr in the Galaxy,
their number still should be much smaller than the white dwarf
binaries, and in this section we restrict ourselves to considering
only binary WD.

Substituting $E=E_{orb}\sim {\M}c^2({\M}f)^{2/3}$ into equation
(\ref{omega_R}) we obtain
\begin{equation}
\Omega_{WD}(f)\approx 2\times 10^{-8} \R_{100} (f/10^{-3}
\hbox{Hz})^{2/3}(\widetilde {\M}/M_\odot)^{5/3}( \langle r
\rangle/10\, \hbox{kpc})^{-2}h_{100}^{-2}\,,
\end{equation}
where $\R_{100}=\R /(0.01$ ~yr$^{-1})$ is the galactic
rate of binary WD mergers.

In terms of the characteristic dimensionless amplitude of the noise
background that determines the signal-to-noise ratio when
cross-correlating outputs of two independent interferometers we have
\begin{equation}
h_c(f)\approx 7.5 \times 10^{-20} \R_{100}^{1/2} 
(f/10^{-3} \hbox{Hz})^{-2/3}(\widetilde {\M}/M_\odot)^{5/6}
( \langle r \rangle/10\,\hbox{kpc})^{-1}
\label{h_c}
\end{equation} 

Equation (\ref{h_c}) shows that at high frequencies of interest here
the GW background is fully determined by the galactic rate of binary
WD mergers and is independent of (complicated) details of binary
evolution at lower frequencies (the examples of calculated spectra at
lower frequencies see in Lipunov \& Postnov 1987; Lipunov, Postnov \&
Prokhorov 1987; Hils et al. 1990).

But the real galactic merger rate of close binary WD is unknown.  One
possible way to recover it is searching for close white dwarf
binaries. A recent study (Marsh et al. 1995), revealed a larger
fraction of such systems than had previously been thought. Still, the
statistics of such binaries in the Galaxy remains very poor.

If coalescing binary WD are associated with SN Ia explosions, as
proposed by Iben \& Tutukov (1984) and further investigated by many
authors (for a recent review of SN Ia progenitors see Branch et
al. 1995), their coalescence rate can be constrained using much more
representative SN Ia statistics. Branch et al. (1995) concluded that
coalescing CO-CO binary WD remain the most plausible candidates mostly
contributing to the SN Ia explosions.  The galactic rate of SN Ia is
estimated $4\times 10^{-3}$ per year (Tamman et al.  1994; van den
Bergh and McClure 1994), which is close to the calculated rate of
CO-CO coalescences ($\sim (1-3)\times 10^{-3}$). The coalescence rate
for He-CO WD and He-He WD (other possible progenitors of SN Ia) falls
ten times short of that for CO-CO WD (Branch et al. 1995).  As SN Ia
explosions may well be triggered by other mechanisms, we conclude that
the observed SN Ia rate provides a secure {\it upper limit} to the
double WD merger rate regardless of the evolutionary considerations.

The upper limit (\ref{h_c}) is plotted in Fig. 1 for different rates
of binary WD mergers $\R_{100}=1, 1/3, 1/10, 1/30$ assuming the chirp
mass ${\M}\approx 0.52 M_\odot$ (as for two CO white dwarfs with equal
masses $M_1=M_2=0.6 M_\odot$).  These lines intersect the proposed
LISA rms sensitivity at 
$
f>f_{lim}\approx 0.03-0.07 \hbox{Hz}\,.
\label{lim}
$ This means that at frequencies higher than $0.07$ Hz no continuous
GW backgrounds of galactic origin are presently known to contribute
above the rms-level of LISA space laser interferometer.  The
contribution from extragalactic binaries is still lower regardless of
the poorly known binary WD merging rate (at least in the limit of no
strong source evolution with $z$). Other possible sources could be
extragalactic massive BH binary systems (e.g. Hils and Bender
1995). Their number in the Universe can be fairly high (e.g. Rees
1997), but no reliable estimates of their contribution are available
at present. The lower limit (\ref{lim}) is already close to the LISA
sensitivity limit at 0.1 Hz, but we stress that the assumptions used
in its derivation are upper limits, so the actual frequency beyond
which no binary stochastic backgrounds contribute may be three times
lower. This precise limit depends on the details of binary WD
formation and evolution which are still poorly known.

Fig. 1 demonstrates that the calculated GW background intersects LISA
sensitivity curve at frequencies $\sim 0.05$ Hz, and Bender et al's
curve at even lower frequencies $\sim 0.01$ Hz. The latter is
probably due to Bender et al's curve being derived from observational
estimate of double WD galactic density in the solar neighborhoods; we
stress once more that once formed, the binary WD will evolve until the
less massive companion fills its Roche lobe; unless the mass ratio is
sufficiently far from one (cf. Webbink 1984), the merger should
occur. Therefore Bender et al's curve provides a secure {\it lower
limit} to the galactic binary stochastic GW background.  

Presently, we cannot rule out the high galactic double WD merger rate
(1/300-1/1000 yr$^{-1}$), and therefore can consider $f_{lim}$ to lie
within the frequency range $0.01-0.07$ Hz.  We conclude that no GW
background of galactic origin above this frequencies should contribute
at the rms-noise level of LISA interferometer, and hence the detection
of an isotropic stochastic signal at frequencies $0.03-0.1$ Hz with an
appreciable signal-to-noise level (which possibly may be done using
one interferometer) would strongly indicate its cosmological origin.
To be detectable by LISA, the power of relic GW background should be
$\Omega_{GW}h_{100}^2>10^{-8}$ in this frequency range.

\begin{figure}
\centerline{
\epsfxsize=0.6\hsize
\epsfbox{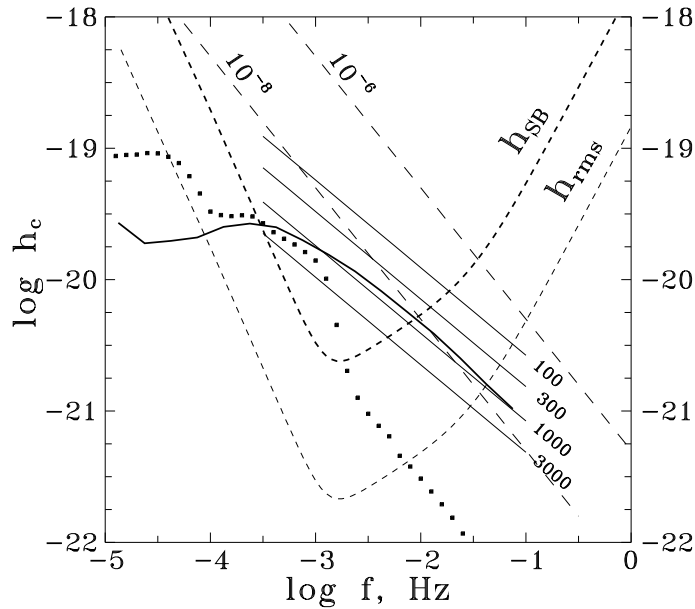}}
\caption{
Galactic binary GW background $h_c$ as given by Bender (1996)
(filled quadrangles) and calculated for a model spiral galaxy
with the total stellar mass $10^{11}$ M$_\odot$ (the solid curve).
Average photometric distance 7.9 kpc
is assumed. Thin straight lines marked with 100, 300, 1000, 3000 are the
analytical upper limit (eq. [\protect\ref{h_c}]) for binary WD merger rates
1/100, 1/300, 1/1000, and 1/3000 yr$^{-1}$ in a model spiral galaxy,
respectively, assuming \protect{$\M=0.52$} M$_\odot$.  
Straight dashed lines labeled by $10^{-8}$, $10^{-6}$ show GW
backgrounds corresponding to constant $\Omega_{GW}$. The proposed LISA rms
noise level ($h_{rms}$) and sensitivity to bursts
$h_{SB}=5\protect\sqrt{5} h_{rms}$ are also reproduced (cf. Thorne 1995;
Fig. 14).}
\end{figure}

\end{document}